\documentclass[onecolumn,superscriptaddress]{revtex4}
\usepackage{hyperref}
\usepackage{graphicx}
\usepackage{multirow}
\usepackage{amsmath,amssymb,latexsym,MnSymbol}
\usepackage{longtable}

\begin{document}

\title{Move-by-move dynamics of the advantage in chess matches reveals population-level learning of the game}

\author{H. V. Ribeiro}\email{hvr@dfi.uem.br}
\affiliation{Departamento de F\'isica and National Institute of Science and Technology for
Complex Systems, Universidade Estadual de Maring\'a, Maring\'a, PR 87020, Brazil}
\affiliation{Department of Chemical and Biological Engineering, Northwestern University, Evanston, IL 60208, USA}
\author{R. S. Mendes}
\author{E. K. Lenzi}
\affiliation{Departamento de F\'isica and National Institute of Science and Technology for
Complex Systems, Universidade Estadual de Maring\'a, Maring\'a, PR 87020, Brazil}
\author{M. del Castillo-Mussot}
\affiliation{Departamento de Estado S\'olido, Instituto de F\'isica, Universidad Nacional Aut\'onoma de M\'exico, Distrito Federal, M\'exico}
\author{L. A. N. Amaral}\email{amaral@northwestern.edu}
\affiliation{Department of Chemical and Biological Engineering, Northwestern University, Evanston, IL 60208, USA}
\affiliation{Northwestern Institute on Complex Systems (NICO), Northwestern University, Evanston, IL 60208, USA}

\begin{abstract}
The complexity of chess matches has attracted broad interest since its invention. This complexity and
the availability of large number of recorded matches make chess an ideal model systems for the
study of population-level learning of a complex system. We systematically investigate the move-by-move dynamics 
of the white player's advantage from over seventy thousand high level chess matches spanning over 150 years. 
We find that the average advantage of the  white player is positive and that it has been increasing over time.  
Currently, the average advantage of the white player is $\sim$0.17 pawns but it is exponentially approaching a value of 0.23 pawns 
with a characteristic time scale of 67 years.  We also study the diffusion of 
the move dependence of the white player's advantage and find that it is non-Gaussian, 
has long-ranged anti-correlations and that after an initial period with no diffusion 
it becomes super-diffusive.  We find that the duration of  the non-diffusive period, 
corresponding to the opening stage of a match, is increasing in length and exponentially 
approaching a value of 15.6 moves with a characteristic time scale of 130 years.  
We interpret these two trends as a resulting from learning of the features of the game. 
Additionally, we find that the exponent $\alpha$ characterizing the super-diffusive 
regime is increasing toward a value of 1.9, close to the ballistic regime. 
We suggest that this trend is due to the increased broadening of the range 
of abilities of chess players participating in major tournaments.
\end{abstract}

\maketitle

\section*{Introduction}
The study of biological and social complex systems has been the focus of intense
interest for at least three decades~\cite{Amaral}.
Elections~\cite{Fortunato}, popularity~\cite{Ratkiewicz}, population growth~\cite{Rozenfeld}, 
collective motion of birds~\cite{Bialek} and bacteria~\cite{Peruani} are just some examples
of complex systems that physicists have tackled in these pages. 
An aspect rarely studied due to the lack of enough data over a long enough period is the manner
in which agents learn the best strategies to deal with the complexity of the system. For example,
as the number of scientific publication increases, researchers must learn how to
choose which papers to read in depth~\cite{Stringer}; while in earlier times word-of-mouth
or listening to a colleague's talk were reliable strategies, nowadays the journal in which
the study was published or the number of citations have become, in spite of their many caveats, 
indicators that seem to be gaining in popularity.

In order to understand how population-level learning occurs in the ``real-word,'' we study it
here in a model system. Chess is a board game that has fascinated humans ever since its invention
in sixth-century India~\cite{OBrien}. Chess is an extraordinary complex game with $10^{43}$ 
legal positions and $10^{120}$ distinct matches, as roughly estimated
by Shannon~\cite{Shannon}. Recently, Blasius and T\"onjes~\cite{Blasius} have showed
that scale-free distributions naturally emerge in the branching process in the game tree 
of the first game moves in chess. Remarkably, this breadth of possibilities emerges from a small set of well-defined rules.
This marriage of simple rules and complex outcomes has made chess an excellent test bed for 
studying cognitive processes such as learning~\cite{Gobet,Saarilouma}  and also for 
testing artificial intelligence algorithms such as evolutionary algorithms~\cite{Fogel}.

The very best chess players can foresee the development of a match 10--15 moves into the future, 
thus making appropriate decisions based on his/her expectations of what his opponent will do.  
Even though super computers can execute many more calculations and hold much more 
information in a quickly accessible mode, it was not until heuristic rules were developed 
to prune the set of possibilities that computers became able to consistently beat human players. 
Nowadays, even mobile chess programs such as Pocket Fritz\texttrademark~(\url{http://chessbase-shop.com/en/products/pocket_fritz_4})
have a Elo rating~\cite{Elo} of $2938$ which is higher than the current best chess player (Magnus Carlsen with a Elo rating of 2835 --- \url{http://fide.com}).

The ability of many chess engines to accurately evaluate the strength of a position enables 
us to numerically evaluate the move-by-move white player advantage $A(m)$ and to determine 
the evolution of the advantage during the course of a chess match. In this way, we can probe 
the patterns of the game to a degree not before possible and can attempt to uncover 
population-level learning in the historical evolution of chess match dynamics. Here,
we focus on the dynamical aspects of the game by studying the move-by-move dynamics of the 
white player's advantage $A(m)$ from over seventy thousand high level chess matches.

\begin{figure}[!ht]
\centering
\includegraphics[scale=0.64]{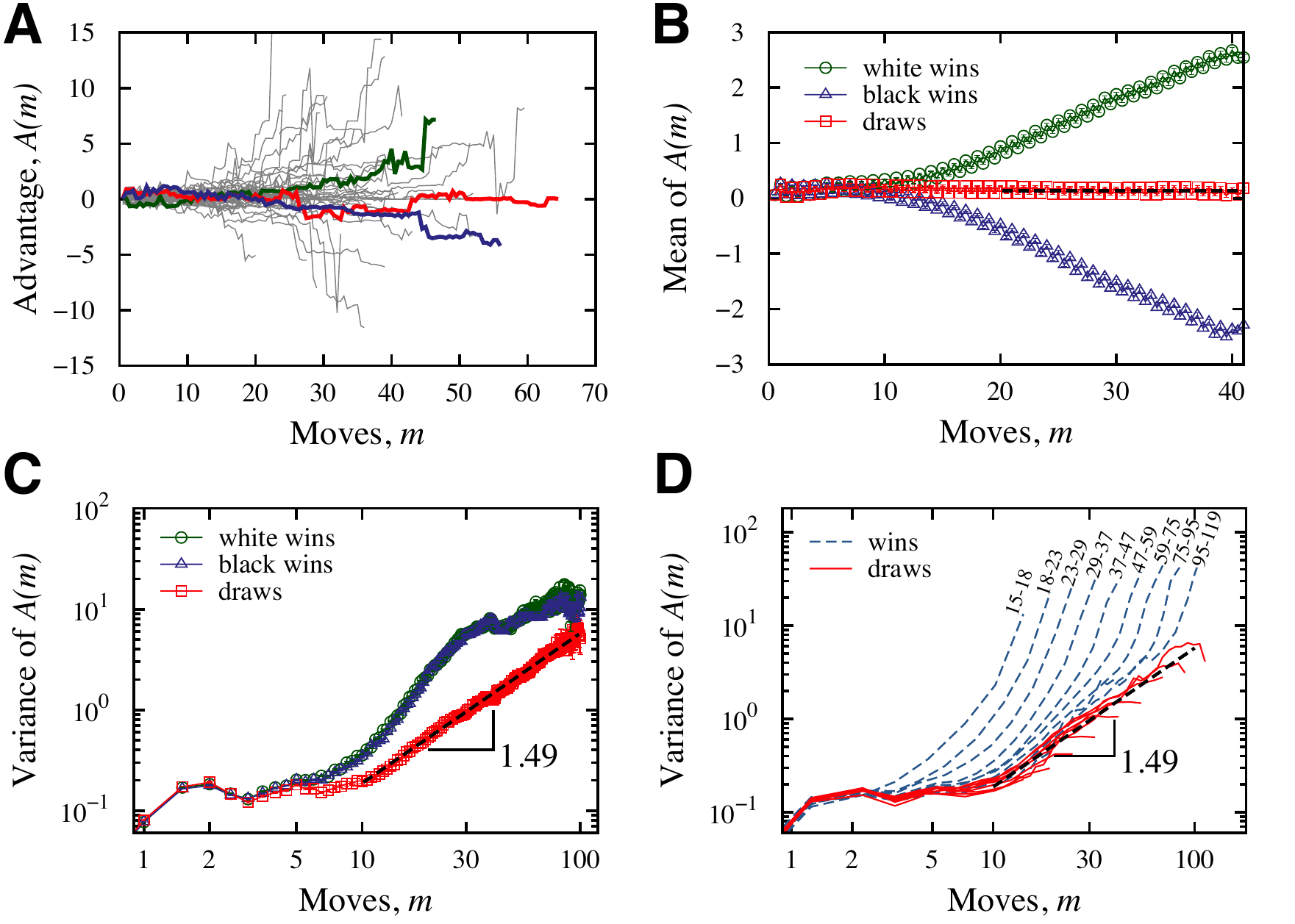}
\caption{{{\bf Diffusive dynamics of white player's advantage.} {(A)} Evolution of the
advantage $A(m)$ for $50$ matches selected at random. We highlight the trajectories
from three World Chess Championship matches: the $6^{\text{th}}$ match between Anand (playing white) and 
Kramnik in 2008 (green line), the $2^{\text{nd}}$ match between Karpov (playing white) and Kasparov in 1985 (red line),
and the $1^{\text{st}}$ match between Spassky (playing white) and Petrosian in 1969 (blue line). 
{(B)} Mean value of the advantage as a function of move number for matches ending in draws (squares), white wins (circles) and black wins (triangles).
Note the systematically alternating values and the initial positive values of these means for all outcomes. 
For white wins, the mean advantage increases with $m$, while for black wins it decreases. For draws, the mean advantage
is approximately a positive constant. We estimated the advantage 
of playing white to be $0.14\pm0.01$ and horizontal dashed line represents this value.
{(C)} Variance of the advantage as a function of move number for matches
ending in draws (squares) and  white wins (circles) and black wins (triangles). Note the very similar profile of the variance for
white and black wins. Note also that there is practically no diffusion for the initial $7-10$ moves, corresponding to the opening period, a very well studied stage of the game. After the opening stage, the trajectories exhibit a faster than diffusive 
spreading. For draws, we find this second regime ($10<m<100$) to be superdiffusive and characterized by
an exponent $\alpha=1.49\pm0.01$, as shown by the dashed line. For wins, the variance presents a more complex behavior. 
For $10 \lesssim m \lesssim40$ the variance increases faster than ballistic (hyper-diffusion), but for 
later stages it displays a behavior similar to that found for draws. {{(D)} Variance of advantage evaluated after grouping the
matches by length and outcome. For draws (continuous lines), the different match lengths do not change the power-law dependence of
the variance. For wins (dashed lines), the variance systematically approaches the profile obtained for draws as the matches 
becomes longer. We further note the existence of a very fast diffusive regime for the latest moves of each grouping.}
}
}\label{fig1}
\end{figure}

We have accessed the portable game notation (PGN) files of 73,444 high level chess matches
made free available by PGN Mentor\texttrademark~({http://www.pgnmentor.com}).
These data span the last two centuries of the chess history and cover the most important worldwide chess tournaments, including the World Championships, Candidate Tournaments, and the Linares Tournaments (see supplementary Table 1). 
White won $33\%$ of these matches, black won $24\%$ and $43\%$ ended up with in a draw. For each of these 73,444 matches, 
we estimated $A(m)$ using the Crafty\texttrademark~\cite{Crafty} chess engine which has an Elo rating of 2950 (see Methods Section A).
The white player advantage $A(m)$ takes into account the differences in the number and the
value of pieces, as well as the advantage related to the placement of pieces. It is usually measured
in units of pawns, meaning that in the absence of other factors, it varies by one 
unit when a pawn (the pieces with lowest value) is captured. A positive value indicates 
that the white player has the advantage and a negative one indicates that the black player
has the advantage. Figure~\ref{fig1}A illustrates the move dependence of $A$ for 50 matches selected at random 
from the data base. Intriguingly, $A(m)$ visually resembles the ``erratic'' movement of diffusive particles.

\section*{Results}
We first determined how the mean value of the advantage depends on the move number $m$ across all matches with the same
outcome (Fig.~\ref{fig1}B). 
We observed an oscillatory behavior around a positive value with a period of $1$ move 
for both match outcomes. This oscillatory behavior
reflects the natural progression of a match, that is, the fact that the players 
alternate moves. Not surprisingly, for matches ending in a draw the average oscillates 
around an almost stable value, while for white wins it increases systematically and for black wins it
decreases systematically.

Figure~\ref{fig1}B suggests an answer to an historical debate among chess players: 
Does playing white yield an advantage?
Some players and theorists argue that because the white player starts the game, 
white has the ``initiative,'' and that black must endeavor to equalize the situation. 
Others argue that playing black is advantageous because white has to reveal the first move. 
Chess experts usually mention that white wins more matches as evidence of
this advantage. However, the winning percentage does not indicate the magnitude of this advantage.
In our analysis, we not only confirm the existence of an advantage in playing white, 
but also estimate its value as $0.14\pm0.01$
by averaging the values of the mean for matches ending in draws.

We next investigated the diffusive behavior by evaluating the dependence of the variance of $A$ on
the move number $m$ (Fig.~\ref{fig1}C).
After grouping the matches by match outcome, 
we observed for all outcomes that there is practically no diffusion during the initial moves. These moves correspond
to the opening period of the match, a stage very well studied and for which there are recognized 
sequences of moves that result in balanced positions. After this initial stage, 
the variance exhibits an anomalous 
diffusive spreading. For matches ending in a draw, we found a super-diffusive regime
($10<m<100$) that is  described by a power law with an exponent $\alpha=1.49\pm0.01$.
We note the very similar profile of the variance of matches ending in white or black
wins. 

Matches ending in a win display an hyper-diffusive regime ($\alpha>2$) --- a signature of 
nonlinearity and out-of-equilibrium systems~\cite{Siegle}. In fact, the behavior for matches ending in wins
is quite complex and dependent on the match length (Fig.~\ref{fig1}D). While grouping the matches by length does not 
change the variance profile of draws, for wins it reveals a very interesting pattern: As the match length increases
the variance profile become similar to the profile of draws, with the only differences occurring in the last moves.
This result thus suggests that the behavior of the advantage of matches ending in a win is very similar to a draw. The main 
difference occurs in last few moves where an avalanche-like effect makes the advantage undergo large fluctuations.

\subsection*{Historical Trends}
\begin{figure}[!t]
\centering
\includegraphics[scale=0.44]{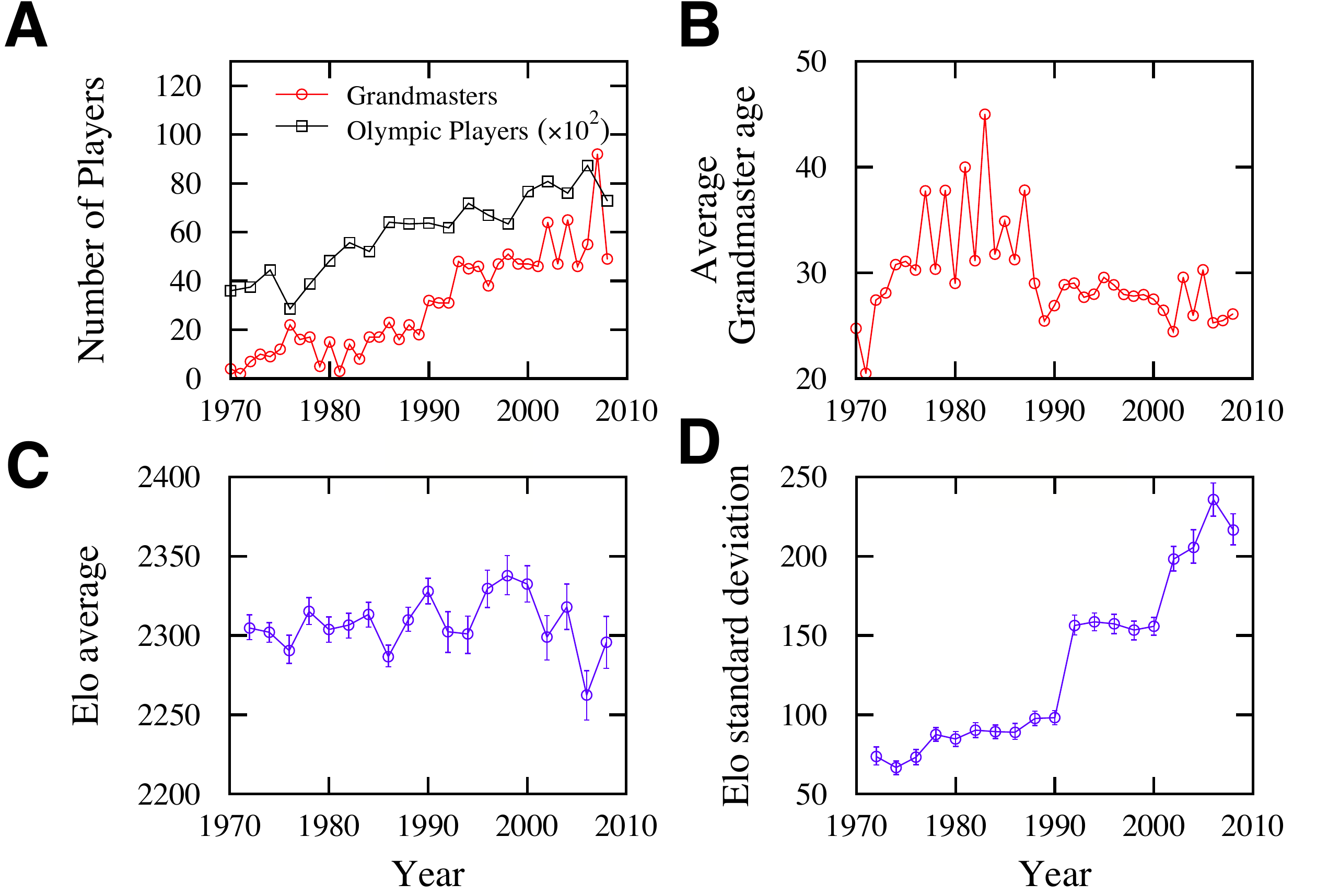}
\caption{{{\bf Historical changes in chess player demographics.} 
{(A)} Number of new Chess Grandmaster awarded annually by the world chess organization (\url{http://fide.com}) 
and the number of players who have participated in the Chess Olympiad (\url{http://www.olimpbase.org}) since 1970.
Note the increasing trends in these quantities. {(B)} Average players' age when receiving
the Grandmaster title. {(C)} Average Elo rating and {(D)} standard deviation
of the of Elo rating of players who have participated in the Chess Olympiad. Note the nearly constant value of the average,
while the standard deviation has increased dramatically.
}
}\label{fig22}
\end{figure}

Chess rules have been stable since the 19th century. This stability increased the game popularity (Fig.~\ref{fig22}A)
and enabled players to work toward improving their skill. A consequence of these efforts is the increasing number of Grandmasters --- the highest title
that a player can attain --- and the decreasing average player's age for receiving this
honor (Figs.~\ref{fig22}A and \ref{fig22}B). Intriguingly, the average player's fitness (measured as the Elo rating~\cite{Elo}) 
in Olympic tournaments has remained almost constant, while the standard deviation of the player's fitness has increased fivefold
(Figs.~\ref{fig22}C and \ref{fig22}D). These historical trends prompt the question of whether there has been a change in the diffusive
behavior of the match dynamics over  the last 150 years. 

\begin{figure}[!t]
\centering
\includegraphics[scale=0.6]{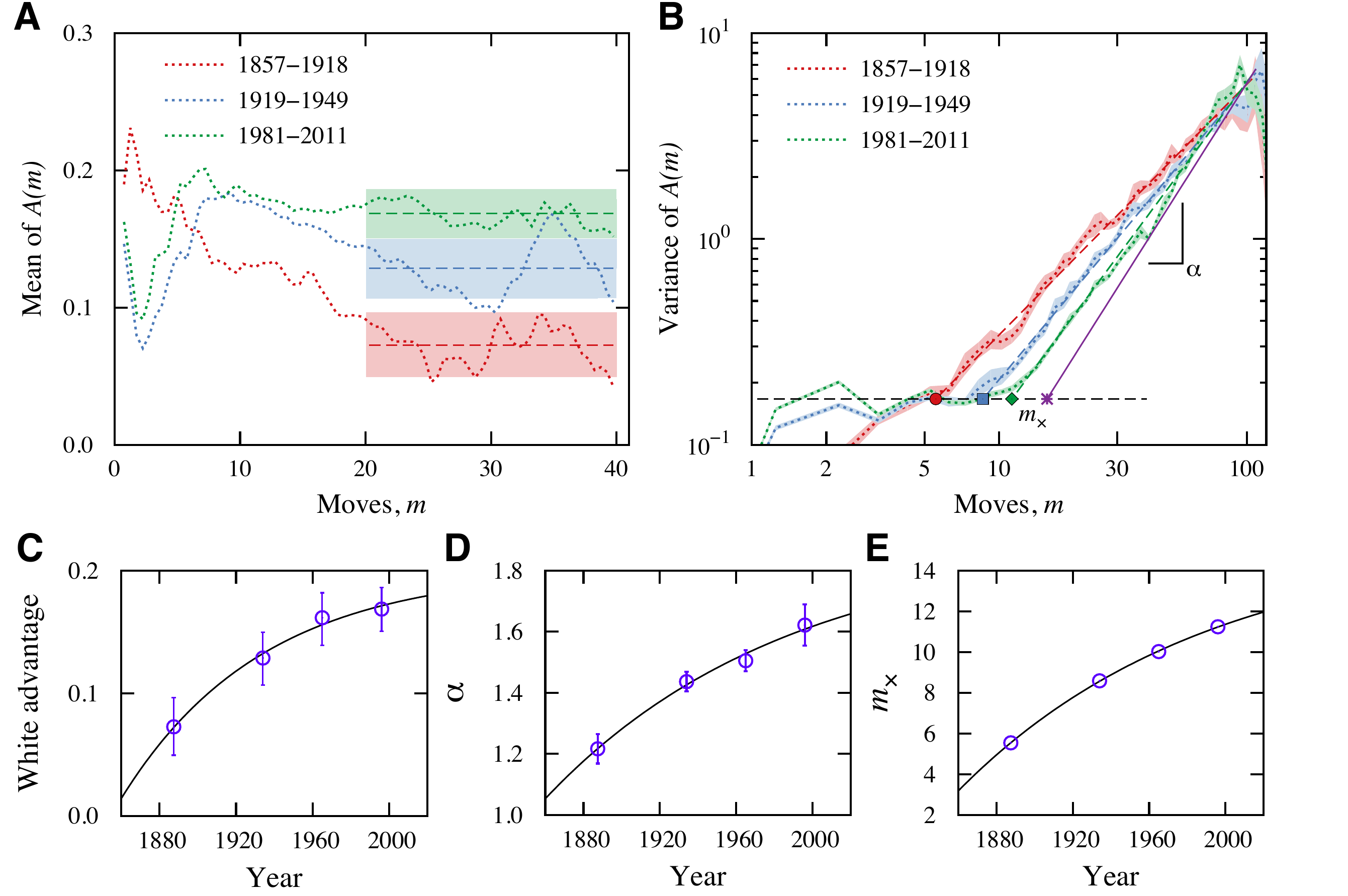}
\caption{{{\bf Historical trends in the dynamics of highest level chess matches.} {(A)}  Mean value of the advantage {of matches ending a draw}
for three time periods. 
These curves were smoothed by using moving averaging over windows of size 2. The horizontal
lines are the averaged values of the mean for $20<m<40$ and the shaded regions are $95\%$ confidence intervals for these averaged values.
{(B)} Variance of the advantage {of matches ending a draw} for three time periods. The shaded regions are $95\%$ confidence intervals for the variance and the colored dashed lines indicate power law fits to each data set. The horizontal dashed line represents the average variance for the most recent data set and for $1<m<10$. Note the systematic increase of $\alpha$ and of the number of moves in the opening. The symbols on this line indicate the values of $m_\times$, the number of moves at which the diffusion of the advantage changes behavior. The rightmost symbol represent the extrapolated maximum value $m_\times=15.6\pm0.6$.
{(C)} Time evolution of the white player advantage for matches ending in draws. The solid
line represents an exponential approach to an asymptotic value. The estimated plateau value is 
$0.23\pm0.01$ pawns and the characteristic time is $67.0\pm0.1$ years.
Time evolution of {(D)} the exponent $\alpha$ and {(E)} the crossover move $m_\times$. The solid lines are fits to exponential approaches to the asymptotic values $\alpha=1.9\pm0.1$
and $m_\times=15.6\pm0.6$. The estimated characteristic times for convergence are $128\pm9$ years for the diffusive exponent
and $130\pm12$ years for the crossover move. Based on the conjecture that $\alpha$ and $m_\times$ are approaching limiting values,
we plotted a continuous line in Fig~3B to represent this limiting regime.
}}\label{fig2}
\end{figure}

To answer this question, we investigated the evolution of the profile of the mean advantage for different periods (Fig.~\ref{fig2}A). For easier visualization,
we applied a moving averaging with window size two to the mean values of $A(m)$. 
The horizontal lines show the average values of the means for $20<m<40$ and the shaded areas are $95\%$ confidence intervals obtained 
via bootstrapping. The average values are significantly different, showing that the
baseline white player advantage has increased over  the last 150 years. We found that this increase is well
described by an exponential approach with a characteristic time of $67.0\pm0.1$ years to an asymptotic value 
of $0.23\pm0.01$ pawns (Fig.~\ref{fig2}C). Our results thus suggest that chess players are learning how to maximize
the advantage of playing white and that this advantage is bounded.

Next, we considered the time evolution of the variance for matches ending in draws (Fig.~\ref{fig2}B). Surprisingly, $\alpha$
seems to be approaching a value close to that for a ballistic regime. We found that the exponent $\alpha$
follows an exponential approach with a characteristic time of $128\pm9$ years to the 
asymptote $\alpha=1.9\pm0.1$ (Fig.~\ref{fig2}D). We surmise that this trend is directly connected
to an increase in the typical difference in fitness among players. Specifically, the presence of fitness in a diffusive process 
has been shown to give rise to ballistic diffusion~\cite{Skalski}. For an illustration of how
differences in fitness are related to a ballistic regime ($\alpha=2$), assume that
\begin{equation}
A_i(m+0.5)=A_i(m)+\Phi_i+\eta(m)
\end{equation}
describes the advantage of the white player in a match $i$, where the difference in fitness between two players
is $\Phi_i$ and $\eta(m)$ 
is a Gaussian variable. $\Phi_i>0$ yields a positive drift in $A_i(m)$ thus modeling a match where the white player is better. 
Assuming that the fitness $\Phi_i$ is drawn from a distribution with finite variance $\sigma^2_{\Phi}$,
it follows that
\begin{equation}
\sigma^2(m)\sim \sigma^2_{\Phi}\,m^2 \,.
\end{equation}
Thus, $\alpha=2$. In the case of chess, the diffusive scenario is not determined purely by the fitness of players. 
However, differences in fitness are certainly an essential ingredient and thus Eq.(1)
can provide insight into the data of Fig.~\ref{fig2}D by suggesting that the typical 
difference in skill between players has been increasing.

A striking feature of the results of Fig.~\ref{fig2}B is the drift of the crossover move $m_\times$
at which the power-law regime begins. We observe that $m_\times$ is exponentially approaching an asymptote at $15.6\pm0.6$ moves
with a characteristic time of $130\pm12$ years (Fig.~\ref{fig2}E). Based on the existence of limiting values 
for $\alpha$ and $m_\times$, we plot in Figure~\ref{fig2}B an extrapolated power law to represent the limiting diffusive regime
(continuous line). We have also found that the distributions of the match lengths for wins and draws display exponential 
decays with characteristics lengths of $13.22\pm0.02$ moves for draws and $11.20\pm0.02$ moves for wins.
Moreover, we find that these characteristic lengths have changed over the history of chess. For matches ending in draws, 
we observed a statistically significant growth of approximately $3.0\pm0.7$ moves per century. 
For wins, we find no statistical evidence of growth and the characteristic length can be approximated by a constant 
mean of $11.3\pm0.6$ moves (supplementary Fig. 1).

A question posed by the time evolution of these quantities is whether the observed changes are due to learning by chess 
players over time or due to a secondary factor such as changes in the organization of chess tournaments. In order to 
determine the answer to this question, we analyze the type of tournaments included in the database. 
We find that 88$\%$ of the tournaments in the database use ``round-robin'' pairing (all-play-all) and that there has been 
an increasing tendency to employ this pairing scheme (supplementary Fig. 2). In order to further strengthen our conclusions, 
we analyze the matches in the database obtained by excluding tournaments that do not use round-robin pairing. 
This procedure has the advantage that it reduces the effect of non-randomness sampling. As shown in 
supplementary Fig. 3, this procedure does not change the results of our analyses.

\begin{figure}[!t]
\centering
\includegraphics[scale=0.64]{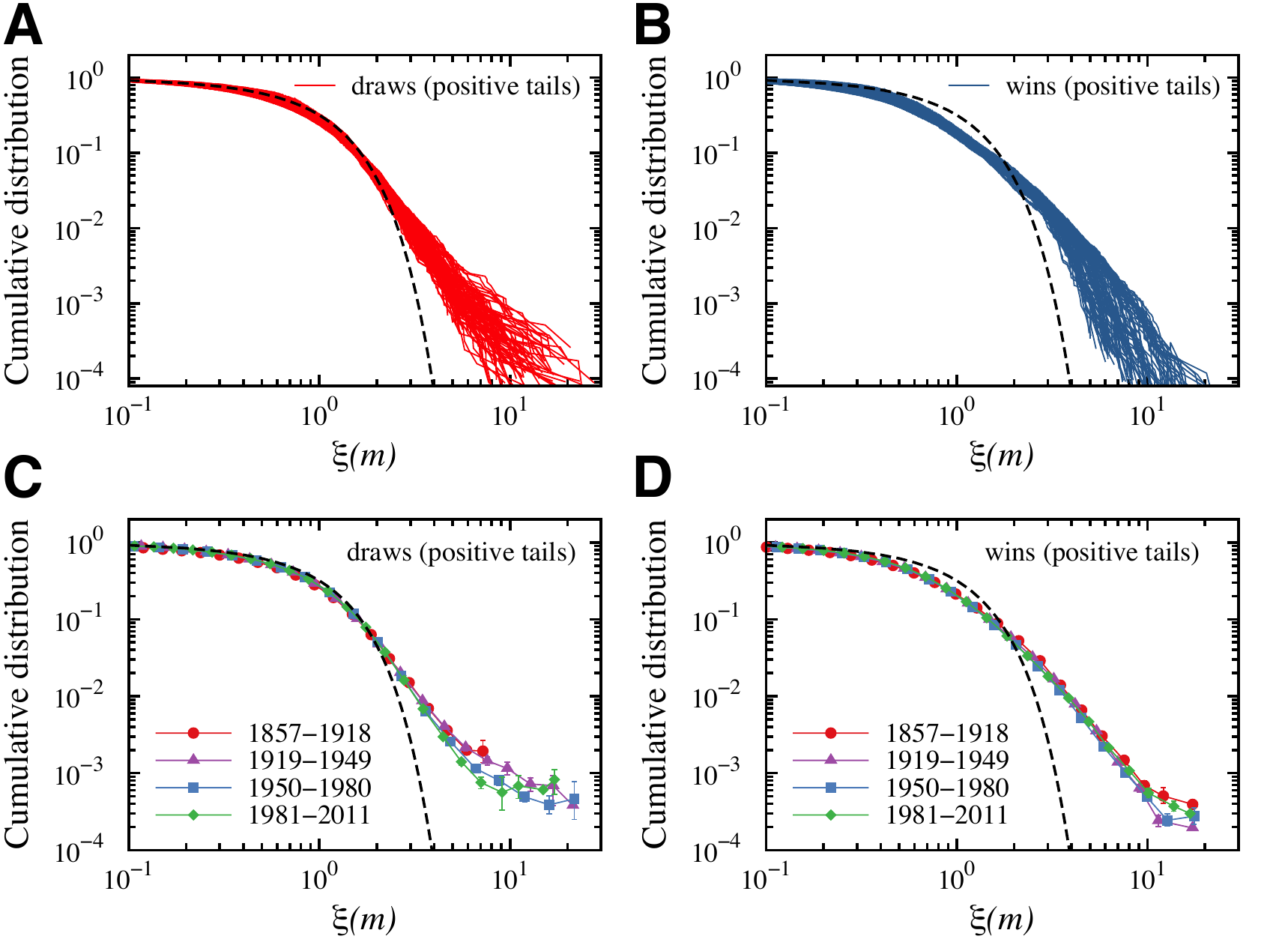}
\caption{{{\bf Scale invariance and non-Gaussian properties of the white player's advantage.}  Positive tails of the cumulative 
distribution function for the normalized advantage $\xi(m)$ for matches ending in {(A)} draws and {(B)} wins. 
Each line in these plots represents a distribution for a different value of $m$ in the range 10 to 70.  By match outcome, the distributions 
for different values of $m$ exhibit a good data collapse with tails that decay slower than a Gaussian distribution (dashed line). Average 
cumulative distribution for matches ending in {(C)} draws and {(D)} wins for four time periods. We estimated the error bars using bootstrapping. 
These data support the hypothesis of scaling, that is, the distributions follow a universal non-Gaussian functional form. The negative tails
present a very similar shape (see supplementary Fig. 4).
}}\label{fig3}
\end{figure}

We next studied the distribution profile of the advantage. We use the normalized advantage 
\begin{equation}
\xi(m)=\frac{A(m)-\langle A(m)\rangle}{\sigma(m)}\,,
\end{equation}
where $\langle A(m)\rangle$ is the mean value of advantage after $m$ moves and $\sigma(m)$ is the standard-deviation.
Figures \ref{fig3}A and \ref{fig3}B show the positive tails of the cumulative distribution of $\xi(m)$  for draws and wins
for $10\leq m\leq70$. We observe the good data collapse, which indicates that the advantages are statistically 
self-similar, since after scaling they follow the same universal distribution. Moreover, Figs.~\ref{fig3}D and \ref{fig3}E 
show that the distribution profile of the normalized advantage is quite stable over the last 150 years. 
These distributions obey a functional form that is significantly different from a Gaussian distribution 
(dashed line in the previous plots). In particular, we observe a more slowly decaying tail, showing the existence of 
large fluctuations even for matches ending in draws.

\begin{figure}[!t]
\centering
\includegraphics[scale=0.64]{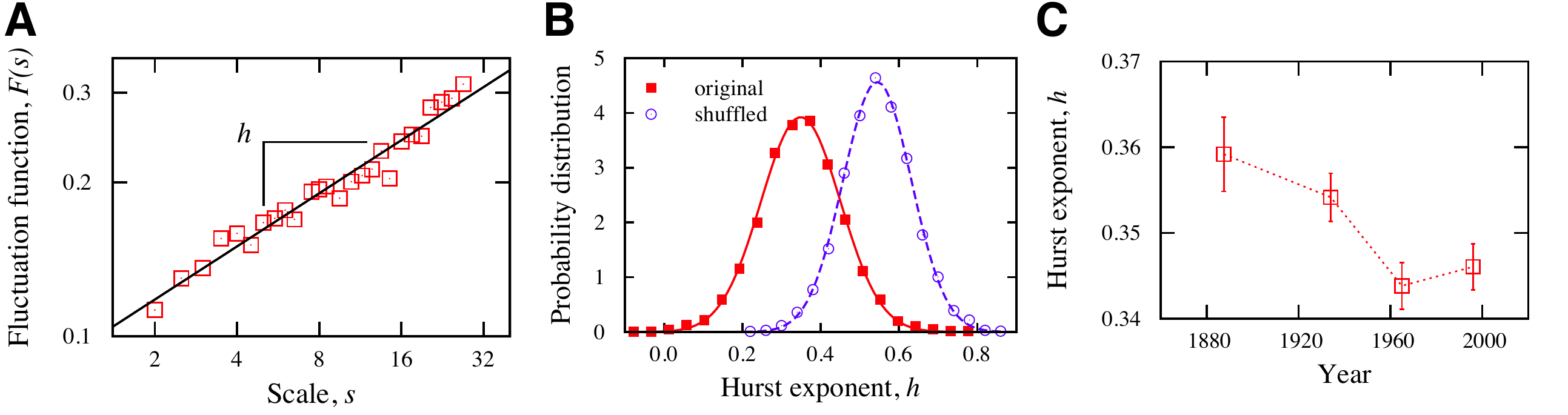}
\caption{{{\bf Long-range correlations in white player's advantage.} {(A)} 
Detrended fluctuation analysis (DFA, see Methods Section B) of white player's advantage increments, that is,
$\Delta A(m) = A(m+0.5)-A(m)$, for a match ended in a draw and selected at random 
from the database. For series with long-range correlations, the relationship between 
the fluctuation function $F(s)$ and the scale $s$ is a power-law where the exponent is the 
Hurst exponent $h$. Thus, in this log-log plot the relationship is approximated by a
straight line with slope equal to $h=0.345$. In general, we find all these relationships
to be well approximated by straight lines with an average Pearson correlation coefficient of $0.892\pm0.002$.
{(B)} Distribution of the estimated Hurst exponent $h$ obtained
using DFA for matches longer than 50 moves that ended in a draw (squares). The continuous line is a Gaussian fit to the 
distribution with mean $0.35$ and standard-deviation $0.1$. Since $h<0.5$, it implies an anti-persistent behavior
(see Fig. \ref{fig1}B). We have also evaluated the distribution of $h$ using the shuffled version of these 
series (circles). For this case, the dashed line is a Gaussian fit to the data with mean $0.54$ and standard-deviation 
$0.09$. Note that the shuffled procedure removed the correlations, confirming the existence of long-range correlations in $A(m)$.
{(C)} Historical changes in the mean Hurst exponent $h$. Note
the significantly small values of $h$ in recent periods, showing that the anti-persistent behavior has increased for more recent matches.}}\label{fig4}
\end{figure}

Another intriguing question is whether there is memory in the evolution of the white player's advantage.
To investigate this hypothesis, we consider the time series of advantage increments 
$\Delta A(m) = A(m+0.5)-A(m)$ for all 5,154 matches ending in a draw that are longer than
$50$ moves. We used detrended fluctuation analysis (DFA, see Methods Section B)
to obtain the Hurst exponent for each match (Fig.~\ref{fig4}A).
We find $h$ distributed around $0.35$ (Fig.~\ref{fig4}B) which indicates the presence of
long-range anti-correlations in the evolution of $A(m)$. A value of $h<0.5$ indicates 
the presence of an anti-persistent behavior, that is, the alternation between large and small values of $\Delta A(m)$ occurs
much more frequently than by chance. This result also agrees with the oscillating behavior of 
the mean advantage (Fig.~\ref{fig1}B). We also find that the Hurst exponent $h$ has evolved over
time (Fig.~\ref{fig4}C). In particular, we note that the anti-persistent behavior has statistically
increased for the recent two periods, indicating that the alternating behavior has intensified in
this period. We have found a very similar behavior for matches ending in wins after removing
the last few moves in the match (supplementary Fig. 5).

\section*{Discussion}
We have characterized the advantage dynamics of chess matches as a self-similar, 
super-diffusive and long-ranged-memory process. Our investigation provides insights
into the complex process of creating and disseminating knowledge of a complex system 
at the population-level. By studying 150 years of high level chess,
we presented evidence that the dynamics of a chess have evolved over 
time in such a way that it appears to be approaching a steady-state. 
The baseline advantage of the white player, the cross-over move $m_\times$, and the 
diffusive exponent $\alpha$ are exponentially approaching asymptotes with different 
characteristic times. We hypothesized that the evolution of $\alpha$ are closely related 
to an increase in the difference of fitness among players, while the evolution of the 
baseline advantage of white player indicates that players are learning better ways to 
explore this advantage. The increase in the cross-over move $m_\times$ suggest that
the opening stage of a match is becoming longer which may also be related to a collective 
learning process. As discussed earlier, hypothesized historical changes in pairing scheme 
during tournaments cannot explain these findings.

\clearpage
\section*{Methods}

\subsection*{Estimating $A(m)$}\label{Estimating_A}
The core of a chess program is called the chess engine. The chess engine
is responsible for finding the best moves given a
particular arrangement of pieces on the board. In order to find the
best moves, the chess engine enumerates and evaluates a huge number of
possible sequences of moves. The evaluation of these possible moves is
made by optimizing a function that usually defines the white player's advantage. 
The way that the function is defined varies from engine to engine, but some key aspects, such as the difference of pondered 
number of pieces, are always present. Other theoretical aspects of chess 
such as mobility, king safety, and center control are also typically considered
in a heuristic manner. A simple example is the definition used for 
the GNU Chess program in 1987 (see {http://alumni.imsa.edu/$\sim$stendahl/comp/txt/gnuchess.txt}).
There are tournaments between these programs aiming to compare the strength of different
engines. The results we present were all obtained using the Crafty\texttrademark~ engine~\cite{Crafty}. 
This is a free program that is ranked 24th in the Computer Chess Rating Lists 
(CCRL - {http://www.computerchess.org.uk/ccrl}). We have also compared the results of 
subsets of our database with other engines, and the estimate of the white player advantage proved robust against those changes.

\subsection*{DFA}\label{DFA}
DFA consists of four steps~\cite{Peng,Kantelhardt}:
\begin{enumerate}
\item[$i)$] We define the profile
$$Y(i)=\sum_{k=1}^{i} \Delta A(m) - \langle \Delta A(m) \rangle\,;$$
\item[$ii)$] We cut $Y(i)$ into $N_s=N/s$ non-overlapping segments of size $s$, where $N$ is the length of the series;
\item[$iii)$] For each segment a local polynomial trend (here, we have used linear function) is calculated and subtracted from $Y(i)$, defining
$Y_s(i)=Y(i)-p_\nu(i)$, where $p_\nu(i)$ represents the local trend in the $\nu$-th segment;
\item[$iv)$] We evaluate the fluctuation function $$F(s)=[\frac{1}{N_s}\sum_{\nu=1}^{N_s} \langle Y_s(i)^2\rangle_\nu]^{1/2}\,,$$
where $\langle Y_s(i)^2\rangle_\nu$ is mean square value of $Y_s(i)$ over the data in the $\nu$-th segment.
\end{enumerate}
If $A(m)$ is self-similar, the fluctuation function $F(s)$ displays a power-law dependence on the time scale $s$, that is, $F(s)\sim s^h$ , where $h$ is the Hurst exponent.

\clearpage

\setcounter{figure}{0}
\makeatletter 
\renewcommand{\thefigure}{S\@arabic\c@figure}
\renewcommand{\thetable}{S\@arabic\c@table}

\section*{\Large Supporting information}

\begin{longtable}{|l|l|}
\caption{{\bf Full description of our chess database.} This table show all the tournaments that comprise our data base.
The PGN files are free available at \url{http://www.pgnmentor.com/files.html}. Specifically, the files we have used are those
grouped under sections ``Tournaments'', ``Candidates and Interzonals'' and ``World Championships''.
}\label{SI-T1} \\

\hline 
\textbf{Tournament} & \textbf{Years} \\ 
\hline 
\endfirsthead

\multicolumn{2}{l}%
{{\bfseries \tablename\ \thetable{} -- continued from previous page}} \\
\hline \textbf{Open} & \textbf{Years} \\ \hline \endhead

\hline \multicolumn{2}{r}{{Continued on next page}} \\ 
\endfoot

\hline \hline
\endlastfoot
\hline \multicolumn{2}{|c|}{{World Championships}} \\ \hline
FIDE Championship & 1996,1998-2000,2002,2004-2008,2010\\
PCA Championship & 1993,1995 \\
World Championship & 1886,1889,1890,1892,1894,1896,1907-1910,1921,1927,1929,1934,1935,\\
& 1937,1948,1951,1954,1957,1958,1960,1961,1963,1966,1969,1972,1978,\\ 
& 1981,1985,1987,1990,1993\\

\hline \multicolumn{2}{|c|}{{Candidates and Interzonals}} \\ \hline
Candidates & 1950,1953,1959,1962,1965,1968,1971,1974,1980,1983,1985,1990,1994\\
Interzonals & 1948,1952,1955,1958,1962,1964,1967,1970,1973,1976,1979,1982,1985,\\
& 1987,1990,1993\\
WCC Qualifier & 1998,2002,2007,2009\\
PCA Candidates & 1994 \\
PCA Qualifier & 1993 \\
World Cup & 2005 \\
\hline \multicolumn{2}{|c|}{{Open Tournaments}} \\ \hline
AVRO & 1938 \\
Aachen & 1868 \\
Altona & 1869,1872 \\
Amsterdam & 1889,1920,1936,1976-1981,1985,1987,1988,1991,1993-1996 \\
Bad & 1977 \\
BadElster & 1937-1939 \\
BadHarzburg & 1938,1939 \\
BadKissingen & 1928,1980,1981 \\
BadNauheim & 1935-1937 \\
BadNiendorf & 1927 \\
BadOeynhausen & 1922 \\
BadPistyan & 1912,1922 \\
Baden & 1870,1925,1980 \\
Barcelona & 1929,1935,1989 \\
Barmen & 1869,1905 \\
Belfort & 1988 \\
Belgrade & 1964,1993,1997 \\
Berlin & 1881,1897,1907,1920,1926 \\
Bermuda & 2005 \\
Bern & 1932 \\
Beverwijk & 1967 \\
Biel & 1992,1997,2004,2006,2007 \\
Bilbao & 2009 \\
Birmingham & 1858 \\
Bled & 1931,1961,1979 \\
Bournemouth & 1939 \\
Bradford & 1888,1889 \\
Breslau & 1889,1912,1925 \\
Bristol & 1861 \\
Brussels & 1986-1988 \\
Bucharest & 1953 \\
Budapest & 1896,1913,1921,1926,1929,1940,1952,2003 \\
Budva & 1967 \\
Buenos Aires & 1939,1944,1960,1970,1980,1994 \\
Bugojno & 1978,1980,1982,1984,1986 \\
Cambridge & 1904 \\
Cannes & 2002 \\
Carlsbad & 1907 \\
Carrasco & 1921,1938 \\
Chicago & 1874,1982 \\
Cleveland & 1871 \\
Coburg & 1904 \\
Cologne & 1877,1898 \\
Copenhagen & 1907,1916,1924,1934 \\
Dallas & 1957 \\
Debrecen & 1925 \\
Dortmund & 1928,1973,1975-1989,1991-2007 \\
DosHermanas & 1991-1997,1999,2001,2003,2005 \\
Dresden & 1892,1926 \\
Duisburg & 1929 \\
Dundee & 1867 \\
Dusseldorf & 1862,1908 \\
Enghien & 2003 \\
Foros & 2007 \\
Frankfurt & 1878,1887,1923,1930 \\
Geneva & 1977 \\
Giessen & 1928 \\
Gijon & 1944,1945 \\
Gothenburg & 1909,1920 \\
Groningen & 1946 \\
Hague & 1928 \\
Hamburg & 1885,1910,1921 \\
Hannover & 1902 \\
Hastings & 1895,1919,1922,1923,1925-1927,1929-1938,1945,1946,1949,1950,1953,\\
 & 1954,1957,1959-1962,1964,1966,1969-2004 \\
Havana & 1913,1962,1963,1965 \\
Heidelberg & 1949 \\
Hilversum & 1973 \\
Hollywood & 1945 \\
Homburg & 1927 \\
Hoogeveen & 2003 \\
Johannesburg & 1979,1981 \\
Karlovy & 1948 \\
Karlsbad & 1911,1923,1929 \\
Kecskemet & 1927 \\
Kemeri & 1937,1939 \\
Kiel & 1893 \\
Kiev & 1903 \\
Krakow & 1940 \\
Kuibyshev & 1942 \\
LakeHopatcong & 1926 \\
LasPalmas & 1973-1978,1980-1982,1991,1993,1994,1996 \\
Leiden & 1970 \\
Leipzig & 1876,1877,1879,1894 \\
Leningrad & 1934,1937,1939 \\
Leon & 1996 \\
Liege & 1930 \\
Linares & 1981,1983,1985,1988-1995,1997-2007 \\
Ljubljana & 1938 \\
Ljubojevic & 1975,1977 \\
Lodz & 1907,1935,1938 \\
London & 1862,1866,1872,1876,1877,1883,1892,1900,1922,1927,1932,1946,1980,\\
& 1982,1984,1986 \\
LosAngeles & 1963 \\
Lugano & 1970 \\
Lviv & 2000 \\
Madrid & 1943,1996,1997,1998 \\
Maehrisch & 1923 \\
Magdeburg & 1927 \\
Manchester & 1857,1890 \\
Manila & 1974, 1975 \\
Mannheim & 1914 \\
MardelPlata & 1928,1934,1936,1942-1957,1959-1962,1965-1972,1976,1979,1981,1982 \\
Margate & 1935,1939 \\
Marienbad & 1925 \\
Meran & 1924 \\
Merano & 1926 \\
Merida & 2000,2001 \\
Milan & 1975 \\
MonteCarlo & 1901-1904,1967 \\
Montecatini & 2000 \\
Montevideo & 1941 \\
Montreal & 1979 \\
Moscow & 1899,1901,1920,1925,1935,1947,1956,1966,1967,1971,1975,1981,1985,\\
& 1992,2005-2007 \\
Munich & 1900,1941,1942,1993 \\
Netanya & 1968,1973 \\
NewYork & 1857,1880,1889,1893,1894,1913,1915,1916,1918,1924,1927,1931,1940,\\ 
& 1951 \\
Nice & 1930 \\
Niksic & 1978,1983 \\
Noordwijk & 1938 \\
Nottingham & 1936 \\
Novgorod & 1994-1997 \\
NoviSad & 1984 \\
Nuremberg & 1883,1896,1906 \\
Oslo & 1984 \\
Ostende & 1905-1907,1937 \\
Palma & 1967,1968,1970,1971 \\
Paris & 1867,1878,1900,1924,1925,1933 \\
Parnu & 1937,1947,1996 \\
Pasadena & 1932 \\
Philadelphia & 1876 \\
Podebrady & 1936 \\
Poikovsky & 2004-2007 \\
Polanica & 1998,2000 \\
Portoroz & 1985 \\
Prague & 1908,1943 \\
Ramsgate & 1929 \\
ReggioEmilia & 1985-1989,1991,1992 \\
Reykjavik & 1987,1988,1991 \\
Riga & 1995 \\
Rogaska & 1929 \\
Rosario & 1939 \\
Rotterdam & 1989 \\
Rovinj & 1970 \\
Salzburg & 1943 \\
SanAntonio & 1972 \\
SanRemo & 1930 \\
SanSebastian & 1911,1912 \\
SantaMonica & 1966 \\
Sarajevo & 1984,1999,2000 \\
Scarborough & 1930 \\
Semmering & 1926 \\
Skelleftea & 1989 \\
Skopje & 1967 \\
Sliac & 1932 \\
Sochi & 1973,1982 \\
Sofia & 2005,2007 \\
SovietChamp & 1920,1923-1925,1927,1929,1931,1933,1934,1937,1939,1940,1944,\\
& 1945,1947-1953,1955-1981,1983-1991 \\
StLouis & 1904 \\
StPetersburg & 1878,1895,1905,1909,1913 \\
Stepanakert & 2005 \\
Stockholm & 1930 \\
Stuttgart & 1939 \\
Sverdlovsk & 1943 \\
Swinemunde & 1930,1931 \\
Szcawno & 1950 \\
Teeside & 1975 \\
Teplitz & 1922 \\
TerApel & 1997 \\
Tilburg & 1977-1989,1991-1994,1996-1998 \\
Titograd & 1984 \\
Trencianske & 1941,1949 \\
Triberg & 1915,1921 \\
Turin & 1982 \\
Ujpest & 1934 \\
Vienna & 1873,1882,1898,1899,1903,1907,1908,1922,1923,1937,1996 \\
Vilnius & 1909,1912 \\
Vinkovci & 1968 \\
Vrbas & 1980 \\
Waddinxveen & 1979 \\
Warsaw & 1947 \\
WijkaanZee & 1968-2007 \\
Winnipeg & 1967 \\
Zagreb & 1965 \\
Zandvoort & 1936
\end{longtable}

\begin{figure}[!t]
\centering
\includegraphics[scale=0.64]{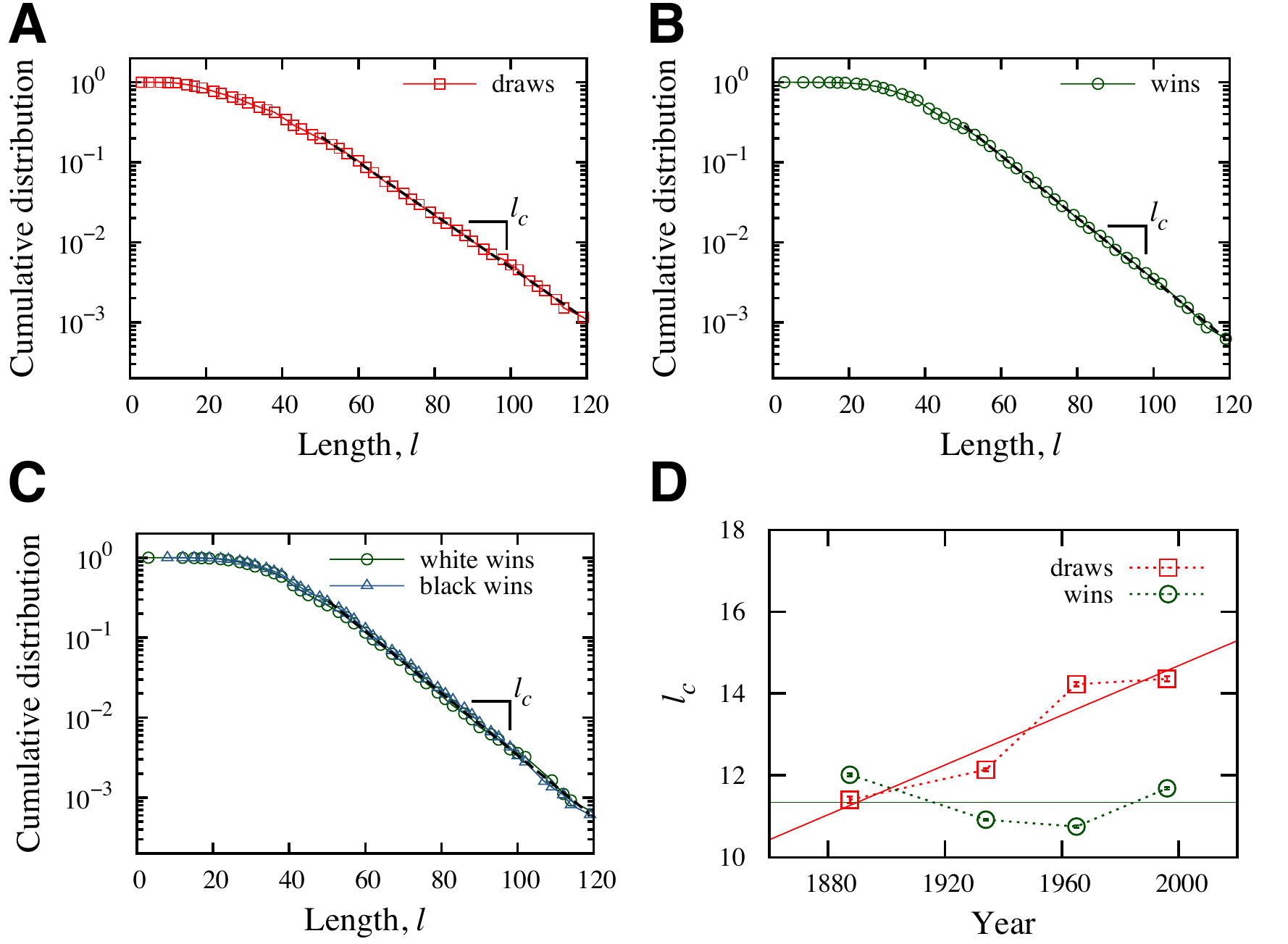}
\caption{{\bf Historical trends in match lengths.} Cumulative distribution function for the match lengths ending in {\bf (A)} draws and wins {\bf (B)}. 
Both distributions display an exponential decay with characteristic lengths $13.22\pm0.02$ for draws and $11.20\pm0.02$ for wins. {\bf (C)}
Cumulative distribution function for the match lengths ending white wins (circles) and black wins (triangles). Note that both distributions
are almost indistinguishable. {\bf (D)} Changes in the characteristic game length $l_c$ over time. For draws (squares), we observe a statistically significant growth of approximately $3.0\pm0.7$ moves per century (red line). For wins (circles), we find that $l_c$ is approximately constant with mean value $11.3\pm0.6$ (green line).}\label{sfig1}
\end{figure}

\begin{figure}[!t]
\centering
\includegraphics[scale=0.64]{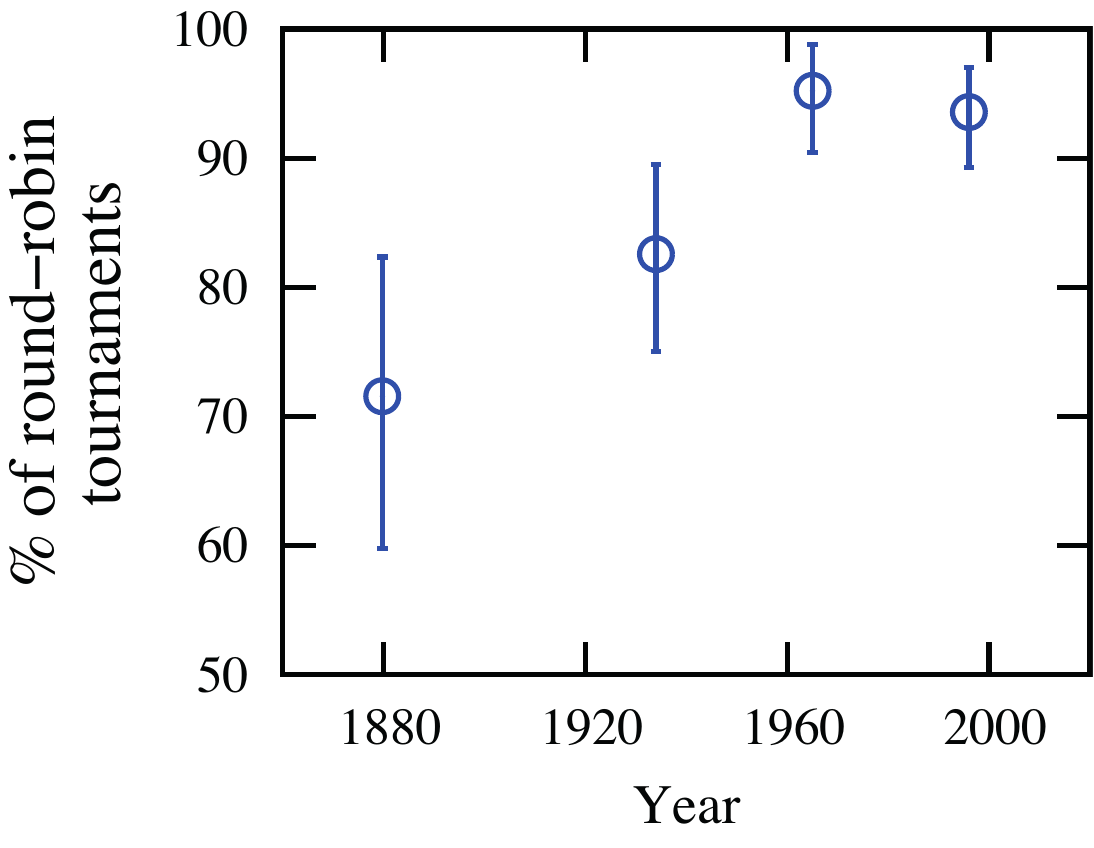}
\caption{{\bf Percentage of tournaments that employ the round-robin (all-play-all) pairing scheme.} 
Note the increase in the fraction of tournaments employing round-robin pairing scheme.}\label{sfig4}
\end{figure}

\begin{figure}[!t]
\centering
\includegraphics[scale=0.5]{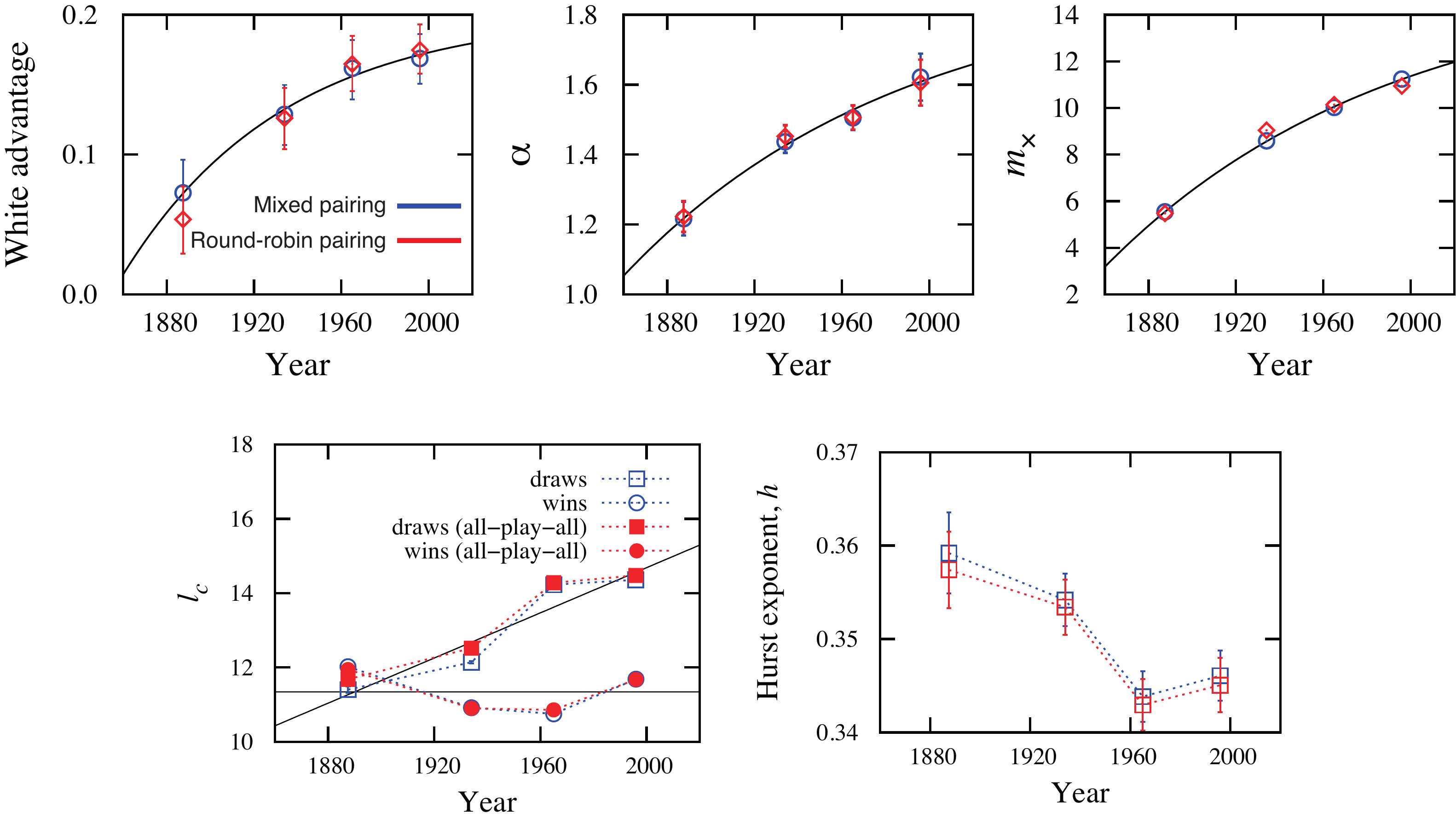}
\caption{{\bf The effect of excluding tournaments using the swiss-pairing scheme on the historical trends reported in Fig. 3.} It is visually apparent that  excluding data from those tournaments does not significantly change our results.  Thus, temporal changes in the pairing schemes used in chess tournaments can not explain our findings.}\label{sfig5}
\end{figure}

\begin{figure}[!t]
\centering
\includegraphics[scale=0.64]{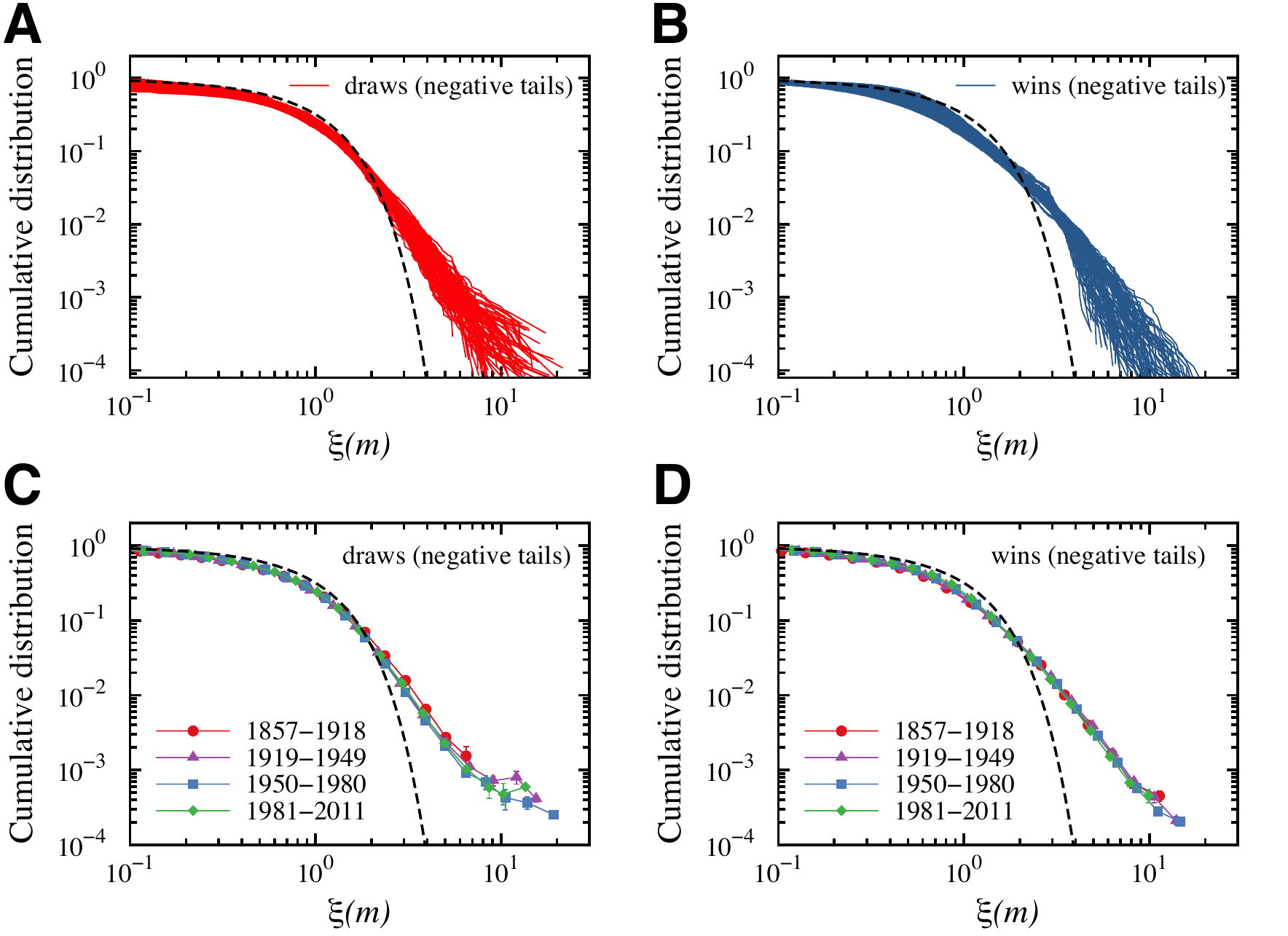}
\caption{{{\bf Scale invariance and non-Gaussian properties of the white player's advantage.} {Negative tails of the cumulative 
distribution function for the normalized advantage $\xi(m)=\frac{A(m)-\langle A(m)\rangle}{\sigma(m)}$ for matches ending in {(A)} draws and {(B)} wins}. 
Each line in these plots represents a distribution for a different value of $m$ in the range 10 to 70. { For match outcome}, the distributions 
for different values of $m$ exhibit a good data collapse with tails that decay slower than a Gaussian distribution (dashed line). Average 
cumulative distribution for matches ending in {(C)} draws and {(D)} wins for four time periods. We estimated the error bars using bootstrapping.}}\label{sfig2}
\end{figure}

\begin{figure}[!t]
\centering
\includegraphics[scale=0.64]{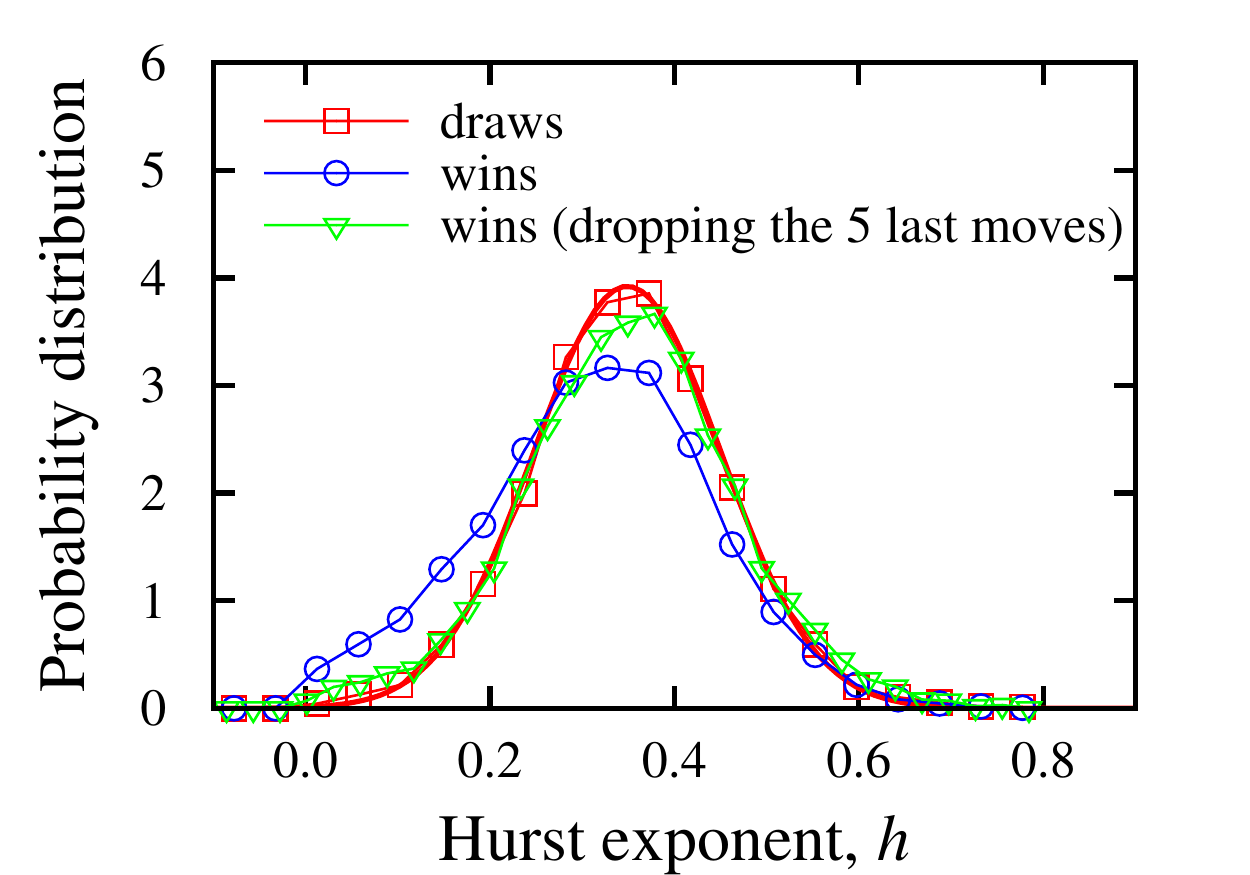}
\caption{{\bf Match outcome and long-range correlations in the white player's advantage.} Distribution of the estimated Hurst exponent $h$ obtained
using DFA for matches longer than 50 moves that ended in draws (squares), wins (circles) and wins after dropping the five last moves of each match. 
The continuous line is a Gaussian fit to the distribution for draws with mean $0.35$ and standard-deviation $0.10$. For wins, the mean
value of $h$ is $0.31$ and the standard-deviation is $0.13$. Note that after dropping the five last moves the distribution of $h$ for wins
becomes very close to distribution obtained for draws. The mean value in this last case is $0.35$ and the standard-deviation is $0.11$.}\label{sfig3}
\end{figure}

\end{document}